\documentclass[a4paper, 11pt]{article}

\usepackage{amsmath}
\usepackage{amsthm}
\usepackage{amssymb}
\usepackage{amscd}
\usepackage{amsfonts}
\usepackage{amsbsy}
\usepackage{setspace}
\usepackage{color}
\usepackage{xcolor}
\usepackage{graphicx}
\usepackage{epstopdf}
\usepackage{calc}
\usepackage{subfigure}
\usepackage{verbatim}
\usepackage{algorithmicx}
\usepackage{setspace}
\usepackage{verbatim}
\usepackage{euscript}
\usepackage{color}
\usepackage{titletoc}
\usepackage{changepage}
\usepackage{framed}
\usepackage{bm}
\colorlet{shadecolor}{blue!10}
\definecolor{light-gray}{gray}{0.9}

\makeatletter
\newcommand{\distas}[1]{\mathbin{\overset{#1}{\kern\z@\sim}}}%
\newsavebox{\mybox}\newsavebox{\mysim}
\newcommand{\distras}[1]{%
  \savebox{\mybox}{\hbox{\kern3pt$\scriptstyle#1$\kern3pt}}%
  \savebox{\mysim}{\hbox{$\sim$}}%
  \mathbin{\overset{#1}{\kern\z@\resizebox{\wd\mybox}{\ht\mysim}{$\sim$}}}%
}
\makeatother

\usepackage[affil-it]{authblk}

\makeatletter
\def\@maketitle{%
  \newpage
  \null
  \vskip 2em
  \begin{center}%
  \let \footnote \thanks
    {\Large\bfseries \@title \par}%
    \vskip 1.5em
    {\normalsize
      \lineskip .5em %
      \begin{tabular}[t]{c}%
         \@author
      \end{tabular} \par}%
    \vskip 1em 
    {\normalsize \@date}%
  \end{center}%
  \par
  \vskip 1.5em} 
\makeatother

\usepackage[english]{babel}
\usepackage[utf8]{inputenc}
\usepackage{amsmath}
\usepackage{amssymb}
\usepackage{graphicx}
\usepackage[colorinlistoftodos]{todonotes}
\usepackage{soul} 

\newcommand{\dP}{{d\!P}}

\newcommand{\argmin}{\operatornamewithlimits{arg\,\,min}}
\newcommand{\ari}{{\rm ARI}}

\providecommand{\keywords}[1]{\textbf{\textit{Keywords: }} #1}

\title{Effects of non-physiological blood pressure artefacts on measures of cerebral autoregulation}
\author{Adam Mahdi$^{a}$, Erica Rutter$^{b}$, Stephen J. Payne$^{a}$}

{
\affil{$^{a\,}$\small Institute of Biomedical Engineering, University of Oxford}

\affil{$^{b\,}$ Department of Mathematics, North Carolina State University} 
}

\date{}

\usepackage{ctable} 

\begin{document}

\maketitle

\onehalfspacing

\begin{abstract}
Cerebral autoregulation refers to  regulation mechanisms that aim to maintain cerebral blood flow approximately constant. It is often assessed by autoregulation index (ARI), which  uses arterial blood pressure and cerebral blood flow velocity time series to produce a ten-scale index of autoregulation performance (0 denoting the absence of and 9 the strongest autoregulation). Unfortunately, data are rarely free from various artefacts. Here, we consider four of the most common non-physiological blood pressure artefacts (saturation, square wave, reduced pulse pressure and impulse) and study their effects on ARI for a range of different artefact sizes.  We show that a sufficiently large saturation and square wave always result in ARI reaching the maximum value of 9. The pulse pressure reduction and impulse artefact lead to a  more diverse behaviour. Finally, we characterized the critical size of artefacts, defined as the minimum artefact size that, on average, leads to a 10\% deviation of ARI. 
\end{abstract}


\section{Introduction}

Cerebral autoregulation (CA) encompasses all the cerebral blood flow regulation mechanisms that maintain cerebral blood flow at an approximately constant level despite changes in arterial blood pressure (ABP). The importance of CA is highlighted by a connection between CA impairment and clinical disorders such as stroke \cite{Dawson2000}, subarachnoid hemorrhage \cite{Giller1990} and head injury \cite{Czosnyka1996}. 

\smallskip

Many different data-driven and physiologically-based approaches have been proposed to assess CA \cite{Payne_CAbook, Panerai08b, Mader2015}. Cerebral autoregulation index (ARI), proposed in 1995 by Tiecks et al. \cite{Tiecks95},  is one of the most popular methods currently used. Given an ABP time series it employs a system of difference equations to predict cerebral blood flow velocity (CBFV), from which a ten-point (0-9) grading index is calculated (0 represents the absence of, and 9 the best autoregulation). 

\smallskip

The reliability of studies involving CA depends heavily on a number of factors including the accuracy of the CA assessment  method and the quality of time series data.  However, the clinical signals are rarely free from various artefacts and the impact on ARI estimates is poorly understood. Previously, Li et al. \cite{Li2009} used a large number of time series to identify the most common non-physiological ABP artefacts. Building on this classification, we consider four artefacts (saturation,  square wave, pulse pressure reduction and impulse) of different magnitudes and embed them in the ABP time series (size 0 corresponds to unperturbed ABP and size 20 corresponds to ABP with the maximum perturbation). Within this framework, we study what effects each of the four artefacts separately can have on ARI. Among other things, we determine the critical artefact, defined here as the size of an artefact that results in an ARI change of 10\% compared to the unperturbed data.

\section{Methods}

\subsection{Data collection and preprocessing}\label{sec:data}
Thirty-six, approximately one-minute, baseline (steady state) ABP and CBFV time series from healthy normotensive subjects are used in the current study. The data collection protocol have previously been discussed in detail in \cite{Lip2000}. The time series were low-pass filtered using zero-phase 4th-order Butterworth filter, in both directions, with a cutoff frequency of $20\,\mathrm{Hz}$.  The beat-to-beat average of ABP and CBFV were calculated for each detected cardiac cycle. The time series were interpolated using a first-order polynomial and subsequently downsampled  at $10\,\mathrm{Hz}$ to produce signals with a uniform time base. Preprocessed ABP and CBFV time series are denoted by $P[k]$ and $V[k]$ and the corresponding mean time series by $\bar P$ and $\bar V$, respectively. 

 \subsection{ARI}\label{sec:ARI}
The computation of ARI follows the original formulation by Tiecks et al. \cite{Tiecks95}. The pressure $P[k]$ is initially normalized: 
\begin{equation}
\dP[k]=\frac{P[k]-\bar P}{\bar P-P_{cr}},
\end{equation}
where $P_{cr}=12\,\mathrm{mmHg}$ is the critical closing pressure.  The method uses a difference model to predict $V[k]$ as follows:
\begin{equation}\label{mod:ARI}
\begin{aligned}
&x_1[k]=x_1[k-1]+\frac{\dP[k]-x_2[k-1]}{fT}\\
&x_2[k]=x_2[k-1]+\frac{x_1[k-1]-2Dx_2[k-1]}{fT}\\
&\hat V[k] \,=\bar V(1+\dP[k]-Kx_2[k]),
\end{aligned}
\end{equation}
where $f$ is the sampling frequency. The three parameters $T, D$ and $K$ are the damping factor, time constant and a parameter reflecting autoregulatory gain, respectively.  Combinations of ten different values of $(T,D,K)$, see \cite{Tiecks95}, are used to generate ten model responses of CBFV, denoted $\hat V_j[k]$, corresponding to various grades of autoregulation, ranging from  0 (absence of autoregulation) to 9 (strongest autoregulation). The difference between the predicted and measured CBFV is computed as  $d_j =  \|(\hat V_j[k] - V[k])/\bar V\|$, where $\|\cdot\|$ is the $l^2$-norm.  The \ari\,\,is  computed as
\begin{equation}\label{ARI}
\ari = \argmin_{s\in\{0,\ldots 9\}} f_{\ari}(s),
\end{equation}
where $f_{\ari}(s)$ is the interpolation by cubic splines of the values $d_j$.

\subsection{Non-physiological blood pressure artefacts}
\label{sec:artdescription}
Here we describe the four common blood pressure artefacts used in the current study, which largely follows the classification given in Li et~al.~\cite{Li2009}.

\medskip

\noindent {\bf Saturation to ABP maximum ($A_{max}$).} 
This artefact is observed as a quick saturation of ABP to some maximum value and it is modelled as
\begin{equation}\label{Amax}
A_{max}(\alpha,L,P_\text{max}) = \tanh\left(\alpha\, \pi\, t\right)(P_\text{max}-P_\text{dias})+P_\text{dias},\qquad t\in[0,L],
\end{equation}
where $t$ is the time of the current value of the artefact and $P_\text{dias}$ is the diastolic blood pressure. The three parameters that govern the shape of the artefact (and their range) are the saturation rate $\alpha\in[0,0.1]$  to the maximal value  $P_\text{max}\in[190,210]$\,mmHg and the length of the duration $L\in[0,5]$\,s.

\medskip

\noindent {\bf  Square wave ($A_{sw}$).}  We assume the square wave artefact is symmetric; the first half being set at maximal blood pressure value and the second half of the square wave at zero. It can be modelled as:
\begin{equation}
A_{sw} (P_\text{max},L)= \left\{
\begin{array}{ll}
P_\text{max}		&\quad t\in[0,L/2]\\
0				&\quad  t\in[L/2,L].
\end{array}\right.
\end{equation}
The shape of the square wave is governed by two paramters: the maximal blood pressure value  $P_\text{max}\in[190, 210]$\,mmHg and the length of the artefact duration $L\in[0,10]$\,s.

\medskip

\noindent {\bf  Pulse pressure  reduction  ($A_{pp}$).}
This artefact appears as a gradual decrease in pulse pressure over time and is usually caused by a thrombus in the arterial line. We simulate the artefact by decreasing the systolic blood pressure linearly over a 45-second window. The slope of the decay is governed by the ratio at the end of the artefact, from 1 (no artifact) to 0.1.

\medskip

\noindent {\bf  Impulse ($A_{\rm imp}$).}
Impulse artefacts appear as a rapid increase in pulse pressure which may last from one to several blood pressure pulses. These are generally caused by motion or mechanical artefacts like crimping of the tube. To model it we define the central lobe of the normalized sinc function
\begin{equation*}
f_L =\left\{
\begin{array}{ll}
 \dfrac{L \sin(2 \pi t/L)}{2 \pi t} &\qquad t\in[-L/2,L/2]\\
 0 					    &\qquad  \text{otherwise}.
\end{array}\right.
\end{equation*}
Note that $f_L$ is continuous and   $f_L=1$ at $t=0$ and parameter $L$ is the width of the normalized central lobe of the $sinc$ function. The impulse artefact can now be simulated by superimposing a scaled $f_L$ on the blood pressure
\begin{equation}\label{Aimp}
A_{\rm imp}(L) =P+ (P_\text{sys}-P_\text{dia})\, f_L, \qquad t\in[-L/2,L/2],
\end{equation}
where $P_\text{sys}$ and $P_\text{dia}$ are the systolic and diastolic pressure, respectively.


\section{Results}

\subsection{Data characterization}
Table~\ref{Table:data} (level 0) gives the mean and standard deviation for the unperturbed (artefact free) ABP signal.  The standard deviation  is understood here as the mean of the standard deviations within each subject along time. 

\subsection{Effects of artefacts on ABP}
\begin{figure}[h!]
\begin{center}
\includegraphics[width=0.75\textwidth]{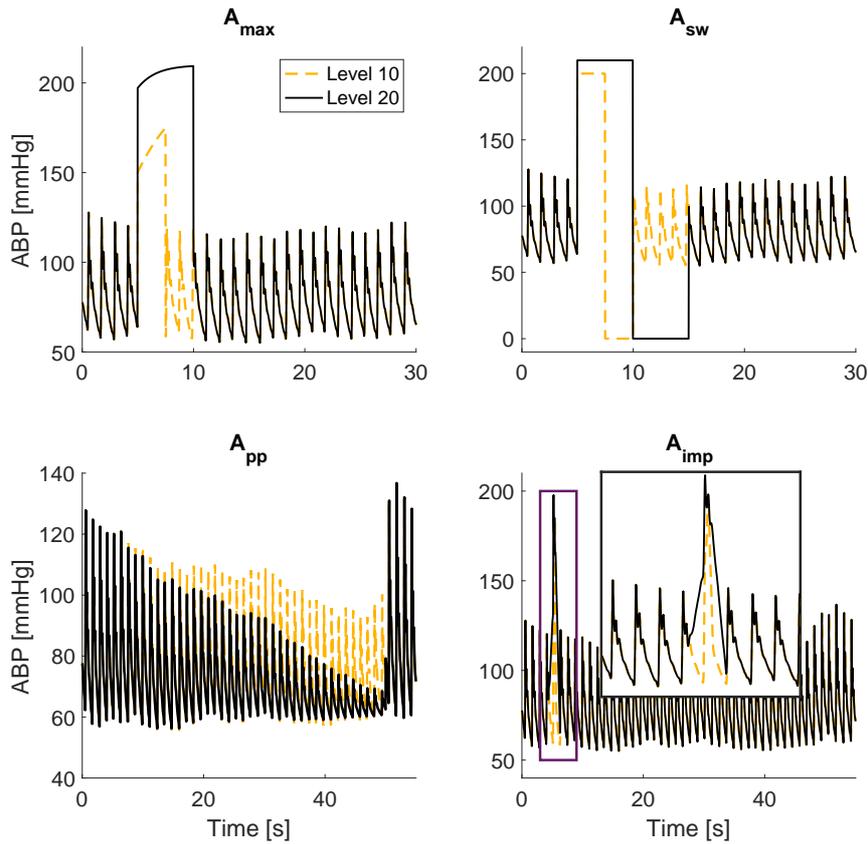}
\caption{{\bf Non-physiological blood pressure artefacts.} The four non-physiological artefacts of size 10 and 20 commonly occurring during ABP measurements.
}\label{Fig:artefacts}
\end{center}
\end{figure}

 Figure~\ref{Fig:artefacts} illustrates each of the four non-physiological artefacts, of size 10 and 20, incorporated into the raw ABP signal, 5 seconds from the beginning of the time series.   Table~\ref{Table:data} (level 10 and 20) shows the corresponding changes in the mean ABP and standard deviation. Since the artefacts are non-physiological, they do not affect the CBFV measurements, so there is no difference in the mean CBFV despite increases in artefact levels.

\subsection{Effects of artefacts on ARI}
\begin{figure}[h!]
\begin{center}
\includegraphics[width=0.75\textwidth]{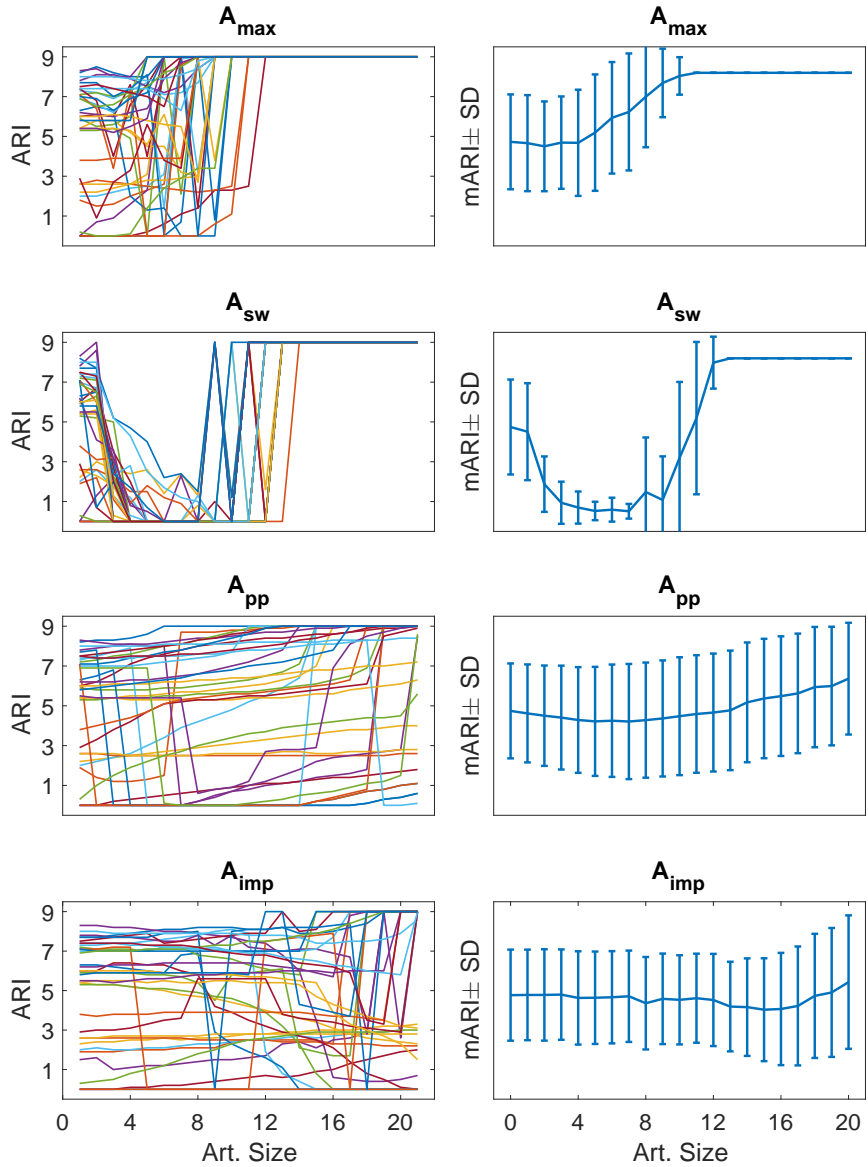}
\caption{{\bf Effects of artefacts on ARI.} {\it Left panel}: The ARI calculated for all normotensive subjects in response to four different types of artefacts.  {\it Right panel}: The mean ARI $\pm$ SD in response to four different levels of artefacts. The size of the artefact are graded on the scale from 0-20 }\label{Fig:all_error}
\end{center}
\end{figure}

Figure~\ref{Fig:all_error} shows the ARI trajectory calculated for all subjects (left panel) and mean ARI and standard deviation (right panel), denoted ${\rm mARI \pm SD}$,  in response to various artefact levels. In the case of saturation and square artefacts, a sufficiently strong perturbation of the signal always results in ARI saturating to the maximum value for all subjects. Additionally, for the square artefact, the ARI dips to almost zero before reaching the maximum value of 9.  The ARI response to the pulse pressure reduction and impulse artefacts, on the other hand, show more moderate changes in the mean ARI.

\subsection{Critical artefact level size}
Table~\ref{tab:maxartlevel} shows the mean and standard deviation of the critical artefacts, i.e. those for which the ARI differs from the initial estimation by 10\%.  Since artefact size is measured somewhat arbitrarily, we also include the corresponding artefacts's parameters (defined in Section \ref{sec:artdescription}). We discard the subjects for which the ARI never changes more than 10\%.

\section{Discussion}

Only a few studies have considered the effects of physiological and non-physiological artefacts  on cerebral blood flow regulation, mainly in the context of the transfer function.    In \cite{EAMES2005} the authors studied the influence of ectopic heart beats, which are naturally occurring episodes. They cause spikes in both the mean ABP and CBFV and can be viewed as a type of physiological artefact. The study showed that replacing ectopic beats by linear interpolation reduced the gain and coherence of the transfer function across the frequency bands.   In \cite{Deegan2011} the authors studied the effects of signal loss in both ABP and CBFV. Their results seem to indicate that the estimates become unreliable with more than 5 seconds of data loss every 50 seconds.  Recently in \cite{Meel-vandenAbeelen2016} the authors investigated the role of three types of artefacts on transfer function: loss of signal, motion artefacts and baseline drifts. Among other things, the study showed that the CA estimates become unreliable when approximately 10\%  of ABP or 8\% of CBFV is lost.

\smallskip

Our results show that  although the four artefacts under consideration strongly affect the ARI values, there were important qualitative differences between them. We note that for a sufficiently large size of the saturation and square wave, ARI always resulted in the maximum value of 9. However,  pulse pressure reduction and impulse exerted a more diverse influence.  For example, for larger size (around 12 and greater)  artefacts the ARI tends to shift upward, but this behaviour is not uniform across individuals. 

\smallskip

The results related to critical artefact, given in Table~\ref{tab:maxartlevel}, corroborate those in Figure~\ref{Fig:all_error}.  Although the critical value is similar for the maximal saturation and the impulse artefact we note that the latter has a larger standard deviation.  There are several potential applications of the critical artefacts. They can be thought of as a signal quality index to flag up  the ARI estimates that are unreliable. Similarly, it can be used in the preprocessing phase to mark the artefacts that must be removed from the signal. 

\smallskip

The current study has several limitations. The approximation of cerebral blood flow by CBFV measured in the MCA is only valid if the diameter of the MCA is constant.  Although currently there are no rigorous  studies showing what is the minimal length needed for the reliable estimation of autoregulation indices, the short time series used in the current study (approximately of 1 minute) might introduce an additional bias to the results.

\section*{Acknowledgement}
AM and SJP acknowledge the support of the EPSRC project EP/K036157/1. The work is supported by the NSF grant DMS 1321794. The authors thank Sang Chalacheva, Kevin O'Keefe, Greg Mader, and Katrina Johnson for simulating conversations.


\begin{thebibliography}{10}

\bibitem{Czosnyka1996}
M~Czosnyka, P~Smielewski, P~Kirkpatrick, DK~Menon, and JD~Pickard.
\newblock Monitoring of cerebral autoregulation in head-injured patients.
\newblock {\em Stroke.}, 27:1829---1834, 1996.

\bibitem{Dawson2000}
SL~Dawson, MJ~Blake, RB~Panerai, and JF~Potter.
\newblock Dynamic but not static cerebral autoregulation is impaired in acute
  ischaemic stroke.
\newblock {\em Cerebrovascular Diseases}, 10(2):126--132, 2000.

\bibitem{Deegan2011}
BM~Deegan, JM~Serrador, K~Nakagawa, E~Jones, FA~Sorond, and G~{\'O}Laighin.
\newblock The effect of blood pressure calibrations and transcranial doppler
  signal loss on transfer function estimates of cerebral autoregulation.
\newblock {\em Medical engineering \& physics}, 33(5):553--562, 2011.

\bibitem{EAMES2005}
PJ~Eames, JF~Potter, and RB~Panerai.
\newblock Assessment of cerebral autoregulation from ectopic heartbeats.
\newblock {\em Clinical science}, 109(1):109--115, 2005.

\bibitem{Giller1990}
CA~Giller.
\newblock The frequency-dependent behavior of cerebral autoregulation.
\newblock {\em Neurosurgery}, 27(3):362--368, 1990.

\bibitem{Li2009}
Q~Li, RG~Mark, and GD~Clifford.
\newblock Artificial arterial blood pressure artifact models and an evaluation
  of a robust blood pressure and heart rate estimator.
\newblock {\em Biomedical {E}ngineering {O}n{L}ine}, 8(1):13, 2009.

\bibitem{Lip2000}
LA~Lipsitz, S~Mukai, J~Hamner, M~Gagnon, and V~Babikian.
\newblock Dynamic regulation of middle cerebral artery blood flow velocity in
  aging and hypertension.
\newblock {\em Stroke}, 31:1897--1903, 2000.

\bibitem{Mader2015}
G~Mader, MS~Olufsen, and A~Mahdi.
\newblock Modeling cerebral blood flow velocity during orthostatic stress.
\newblock {\em Annals of Biomedical Engineering}, 43:1748--1758, 2015.

\bibitem{Meel-vandenAbeelen2016}
ASS Meel-van~den Abeelen, DLK de~Jong, J~Lagro, RB~Panerai, and JAHR Claassen.
\newblock How measurement artifacts affect cerebral autoregulation outcomes: A
  technical note on transfer function analysis.
\newblock {\em Medical engineering \& physics}, 38(5):490--497, 2016.

\bibitem{Panerai08b}
RB~Panerai.
\newblock Cerebral autoregulation: from models to clinical applications.
\newblock {\em Cardiovasc Eng.}, 8:43--59, 2008.

\bibitem{Payne_CAbook}
SJ~Payne.
\newblock {\em Cerebral Autoregulation: Control of Blood Flow in the Brain}.
\newblock Springer, 2016.

\bibitem{Tiecks95}
FP~Tiecks, AM~Lam, R~Aaslid, and DW~Newell.
\newblock Comparison of static and dynamic cerebral autoregulation
  measurements.
\newblock {\em Stroke.}, 26:1014---1019, 1995.

\end{thebibliography}

\section*{Figures}

\clearpage
\section*{Tables}

\begin{table}[h!]
\centering
\begin{tabular}{llccccccc}
\specialrule{.1em}{.05em}{.05em} 
			&          	&	& \multicolumn{2}{c}{Level 0} 	& \multicolumn{2}{c}{Level 10}     & \multicolumn{2}{c}{Level 20}\\
Artefact & { Quantity} 	& Units 		& Mean         	& SD 	      	& Mean 		 	& SD           & Mean 		 	& SD \\
\hline
${\rm A_{max}}$ &ABP 			& mmHg 	    	& 88.43		        	& 11.28	 	      	& 91.29       			& 10.92               & 96.07      			& 10.66 \\
${\rm A_{sw}}$           &ABP 			& mmHg 	    	& 88.43		        	& 11.28	 	      	& 88.20       			& 11.64               & 86.16        			& 10.28 \\
${\rm A_{pp}}$           &ABP 			& mmHg 	    	& 88.43		        	& 11.28	 	      	& 85.28       			& 11.00               & 82.13       			& 10.76 \\
${\rm A_{imp}}$         &ABP 			& mmHg 	    	& 88.43		        	& 11.28	 	      	& 88.71       			& 11.30               & 90.00        			& 11.41 \\
All & CBFV				& cm/s 	    	& 42.73	 	   	& 12.21	 	      	& 42.72		 		& 12.20 	            & 42.74		 		& 12.20 \\
\specialrule{.1em}{.05em}{.05em} 
\end{tabular}
\caption{The main characteristics of the beat-to-beat average ABP and CBFV. The `Mean' is the calculated as the mean across all subjects. Note that Level 0 correspond to the data free from artefacts. 
}\label{Table:data}
\end{table}

\begin{table}[h!]
\centering
\begin{tabular}{llllllll}
\specialrule{.1em}{.05em}{.05em} 
Artefact &  Critical  size  						&  Parameters &&  \\
\hline
${\rm A_{max}}$	& 		$ 5.21\pm 2.1$ 		& L=1.3\,s 				& ${\rm P_{\max}=195.2\,mmHg}$       	&$\alpha= 0.026$ \\
${\rm A_{sw}}$		&		$ 2.75 \pm2.1$ 		& L=1.4\,s       				& ${\rm P_{\max}=192.8\,mmHg}$		&\\
${\rm A_{pp}}$	 	&		$ 8.06 \pm 5.1$ 	&  \text{slope} =  0.64		& 								& \\
${\rm A_{imp}}$		&		$ 11.13 \pm 4.9$ 	& L=0.5\,s       				& 								&\\
\specialrule{.1em}{.05em}{.05em} 
\end{tabular}
\caption{The mean critical artefact size $\pm$ standard deviation and the corresponding parameters that generate it.}
\label{tab:maxartlevel}
\end{table}

\end{document}